\documentclass[useAMS,usenatbib]{mn2ebis}

\usepackage{graphicx}

\def \mref#1{(\ref{#1})} 
\newcommand\sss{\scriptscriptstyle}

\def \jtt {{\rm J}1012$+$5307\ }
\def \jttns {{\rm J}1012$+$5307}
\def \dphil {\Delta\phi_{\sss \rm L}}
\def \dphit {\Delta\phi_{\sss \rm T}}
\def \dbeam {\Delta\phi_b}
\def \dphintr {\Delta\phi_{\rm intr}}
\def \dphi {\Delta\phi}
\def \drl {\Delta r}

\def \phcst {\phi_{\rm cst}}

\def \rcst {r_{\rm cst}}
\def \thcst {\theta_{\rm cst}}
\def \rav {r_{\rm av}}

\def \scst {s_{\rm cst}}

\def \phit {\phi_{\sss \rm T}}

\def \flr  {R_{\sss F}}
\def \rhiof {\rho_{\rm iof}}
\def \rhiofl {\rho_{\rm iof,L}}
\def \rhioft {\rho_{\rm iof,T}}
\def \vaiof {\vec a_{\rm iof}}

\def \ab {a_{\sss B}}
\def \arot {a_{\sss \Omega}}

\def \rhob {\rho_{\sss B}}
\newcommand{\rlc}{R_{\rm lc}}

\newcommand\phf{\phi_f}
\newcommand\phpa{\phi_{\rm PA}}

\newcommand{\thlo}{\theta_{\rm lo}}

  \newcommand\om{(\vec \Omega, \vec \mu)}

\def \rns{R_{\rm NS}}

\title[The rotational asymmetry of pulsar profiles]
{Rotational asymmetry of pulsar profiles}

\author[J. Dyks, G.~A.~E.~Wright, and P. Demorest
]{J. Dyks$^{1}$,
G.~A.~E.~Wright$^{2}$, and P. Demorest$^3$\\
$^{1}$Nicolaus Copernicus Astronomical Center, Toru\'n, Poland\\
$^{2}$Astronomy Centre, University of Sussex, Falmer, BN1 9QJ, UK\\
$^{3}$Department of Astronomy, University of California, Berkeley, CA
94720-3411
}
\begin{document}

\date{Accepted .... Received ...; 
in original form 2008 November 28}

\pagerange{\pageref{firstpage}--\pageref{lastpage}} \pubyear{2002}

\maketitle

\label{firstpage}

\begin{abstract}
We analyse the influence of rotation on shapes of pulse profiles
of fast-rotating (millisecond) pulsars.
Corotation has two opposing effects:
1) the caustic enhancement of the trailing side (TS)
by aberration and retardation (AR), which squeezes the 
emission into a narrower phase interval; 2) the weakening
of the TS caused by the asymmetry of curvature radiation about the dipole axis.
Analysis of the radii of curvature 
of electron trajectories in the inertial observer's frame (IOF) enables 
these two effects to be considered together. 
We demonstrate that for dipolar magnetic field lines
on the TS there exists a `caustic phase' beyond which 
no emission can be observed. 
This phase corresponds to the zero (or minimum) curvature of the IOF 
trajectories and \emph{maximum} bunching of the emission. 
The maximum gradient of polarisation angle (PA) in the S-shaped PA curve
is also associated with the curvature minimum and occurs at exactly the 
same phase.
The asymmetry of trajectory curvature with respect to the dipole axis affects 
the curvature emissivity and the efficiency of pair production, suggesting a 
\emph{minimum} at the caustic phase. Emission over a fixed range of altitudes,
as expected in millisecond pulsars, leads to broad leading profiles and sharp 
peaks with a cutoff phase on the TS. We apply our results to the main pulse 
of the 5 ms pulsar J1012+5307.
\end{abstract}


\section{Introduction}

We analyse two specific issues that are important for the appearance of
pulse-averaged profiles of pulsars:
1) the radial extent of a region emitting at a given radio frequency,
and 2) the asymmetry about the magnetic axis of the curvature of electrons' trajectories
measured in the observer frame. The influence of both of these factors on profile shape
is closely associated with the rotation of the magnetosphere, so they should
be most noticeable for rapidly rotating objects such as the millisecond pulsars (MSPs).
Unlike normal pulsars, MSPs do not exhibit radius-to-frequency mapping (RFM, 
see figs.~16 and 17 in Kramer et al.~1999).
This might possibly be interpreted as near-surface emission with 
negligible radial extent (McConnell et al.~1996).
However, Dyks, Rudak \& Demorest (2009) have shown that
at least some
components in radio pulse profiles are produced when our sightline cuts 
through the curvature emission from radially-extended plasma streams.
In this stream-cut scenario,
the observed location of a component 
is determined by the frequency-independent moment at which our line
of sight crosses the plane of a thin stream. When this happens, the component
becomes simultaneously recorded at all frequencies that are emitted at the
altitude of the cut. There is no RFM,
because positions of components are fixed by the geometry of the cut.

This interpretation requires that a broad range of radio frequencies 
is emitted by 
the stream at a fixed altitude. Such a possibility arises naturally 
in the stream-cut
scenario, because density of plasma in a stream decreases in the transverse 
direction from the stream axis. 
Therefore, regardless of the altitude that is considered, 
a broad range of plasma densities (and plasma frequencies) is available
at different transverse distances from the stream axis.
It is then possible to have a radially-extended emission region (a stream)
which provides a range of plasma frequencies at each altitude
(the `radius-to-frequency' mapping is replaced with 
the `stream-diameter-to-frequency' mapping).
Thus, the radially-extended
emission at a fixed radio frequency is possible without
violating the foundation of the RFM idea, that relates the observed
frequency to the local plasma frequency at the emission 
point.\footnote{Needless to say,
even in the stream the density must also decrease with altitude, so the RFM
is present, but becomes a second-order effect.}

The effect of rotation on a radially-extended source can be expected to make intrinsically symmetric 
pulse profiles appear asymmetric with respect to their centre, changing the phases of their components, their intensities and their widths.
In the following section (\ref{widths}) we
introduce a simple (small-angle) analytical model to determine 
the likely observable effects in the profiles of fast-rotating pulsars with 
emitting regions of finite height.
These caustic effects are well studied {\it numerically} and are routinely employed
to interpret high-energy pulse profiles of pulsars (Morini 1983; Smith 1986;
Romani \& Yadigaroglu 1995; Cheng et al.~2000, Dyks \& Rudak 2003).
Here, however, we focus on the {\it analytical} study of the 
phenomenon, which is feasible for low emission altitudes ($r \la 0.1\rlc$)
or small distances from the dipole axis.
This limitation therefore allows possible applications to the 
radio profiles as well as to specific gamma-ray peaks, eg.~those generated by the 
trailing caustics of the two-pole caustic model.

The results are used to create artificial but instructive pulsar profiles in 
Section (\ref{models}). In Section (\ref{intensity}) we analyse the curvature 
of the particle trajectories and examine the impact of considering curvature 
radiation on model profiles. Then, in Section (\ref{MSP}), we see how far our 
results can model the main pulse profile of J1012+5307. Finally, 
conclusions are drawn in Section (\ref{conc}).


\section{The phase-dependent effects of rotation on pulsar profiles}
\label{widths}
\subsection{Derivation of caustic coordinates}

Our `simplest case' model for an averaged pulse profile is limited 
to the plane of the rotational equator
($\zeta \simeq 90^\circ$, where $\zeta$ is the viewing angle measured
from the rotation axis $\vec \Omega$),
and assumes the orthogonal inclination of dipolar magnetic field 
($\alpha\simeq 90^\circ$). The $\vec B$-field in the CF
is assumed to have the shape of the static dipole.

The radio-emission region is assumed to have the form
of a flaring tube/funnel with negligible
thickness. It is radially extended, but limited
to a single value of the footprint parameter $s$ and it is symmetric
with respect to the $\om$-plane (the fiducial plane).

In the CF the radiation from the radial position $\vec r$
is emitted at the angle
\begin{equation}
\theta_b \simeq \frac{3}{2}s\sqrt{\frac{r}{\rlc}}.
\label{thb}
\end{equation}
with respect to the dipole axis $\vec \mu$,
where the footprint parameter $s=\sin\theta/\sin\thlo$,
$\theta$ is the angle between $\vec r$ and $\vec \mu$,
$\thlo\approx(r/\rlc)^{1/2}$ is the $\theta$-coordinate of the last open 
field lines at the same radial distance $r$.

The radiation emitted on the leading side (LS) of the dipole axis is 
therefore detected at the phase
\begin{equation}
\phi_{\rm L} \simeq -\frac{3}{2}s\sqrt{\frac{r}{\rlc}} - 2\frac{r}{\rlc} +
\phf
\label{phil}
\end{equation}
where the term $2r/\rlc$ accounts for the leftward shift
caused by AR effects, ie.~the aberration (contributing one $r/\rlc$) 
and the retardation (contributing another $r/\rlc$).
The symbol $\phf$ denotes the absolute fiducial phase,
determined
by the detection of a photon emitted from the center of the NS
exactly at the moment when the dipole axis was in the $\vec \Omega$-observer 
plane
(see Dyks 2008 for details).

The radiation on the trailing side (TS) of the dipole axis
is detected at the phase:
\begin{equation}
\phi_{\rm T} \simeq \frac{3}{2}s\sqrt{\frac{r}{\rlc}} - 2\frac{r}{\rlc} +
\phf.
\label{phit}
\end{equation}
The opposing signs of the $r$-dependent terms in eq.~\mref{phit} mean that
on the TS the AR effects tend to cancel 
the effect of the curvature of $\vec B$-field lines, leading to specific caustic effects.
In fact, for arbitrarily large $r$ 
it is easily possible for the TS-radiation
to arrive/be observed on the \emph{leading} side of the profile
(especially if $s$ is small).

Let us consider electrons moving upward along a magnetic field line with parameter $s$
on the TS of the open region. As their altitude increases, their radiation is 
detected at progressively later
phases until, at an altitude of $r_{cst}$, an increment in height of an 
electron's trajectory yields an increase in the linear term 
in eq.~\mref{phit} which is greater than that of the square root term.  
Emission is then detected at the phase $\phcst$, which is `the 
latest-possible detection
phase' for a magnetic fieldline with a given $s$, because further 
increase in $r$ does not produce any advance in detection phase. 
For still larger altitudes,
the radiation becomes detected at phases \emph{earlier} than $\phcst$.
At phases satisfying $\phi \la \phcst$ it is therefore possible
to observe radiation from particles at two different emission altitudes 
on the \emph{same} fieldline 
(because the quadratic eq.~\mref{phit} for $r$ has two real roots). 

For a given $s$ the height $r_{cst}$ and phase $\phcst$ can be found by using eq.~\mref{phit} to solve the equation:
\begin{equation}
\frac{d\phit}{dr} = \frac{1}{\rlc} \left(\frac{3s}{4}\left[\frac{\rlc}{r}
\right]^{1/2} - 2\right)= 0,
\label{phderiv}
\end{equation}
which gives the magnetospheric location of the optimally caustically-enhanced emission:
\begin{equation}
\frac{r_{cst}}{\rlc} \simeq \frac{9}{64}s^2.
\label{rcst}
\end{equation}
Then, from eq.~\mref{phit}:
\begin{equation}
\phcst \simeq \frac{9}{32}s^2 + \phf = 
16^\circ\kern-1.25mm.11 s^2 + \phf.
\label{phcst}
\end{equation}
Thus, for $\alpha\simeq\zeta\simeq90^\circ$ no tangent-to-$\vec B$
radiation emitted within the open fieldline region ($s \le 1$) 
can be detected later than $16^\circ$ after the
fiducial phase, regardless of the emission altitude.

Hereafter the `caustic' coordinates of eq.~\mref{rcst} will be denoted by
$\rcst$ and $\scst=s_{cst}(r)$, and the `caustic enhancement phase' by $\phcst$.

\subsection{Caustic phase compression}

To estimate the deforming effects of aberration and retardation on a pulsar's profile,
let us assume that all radiation is emitted between radial distances $r_1$ and $r_2$ 
(where $r_1 < r_2$), and that $\Delta r < R_{lc}$.  The radial limits will in general refer
to some selected frequency, so strictly we should write $r_{\nu, 1}$ and $r_{\nu, 2}$, but the index $\nu$
will be omitted for simplicity.

From eq.~\mref{phil} the LS radiation emitted between $r_1$ and $r_2$
is therefore spread over the phase range:
\begin{eqnarray}
\dphil& =& \phi_{\rm L}(r_2)-\phi_{\rm L}(r_1) 
\simeq \nonumber\\
& \simeq & -\frac{3}{2}s\left[\sqrt{\frac{r_2}{\rlc}}
-\sqrt{\frac{r_1}{\rlc}}\right]- 2\left[\frac{r_2}{\rlc}-
\frac{r_1}{\rlc}\right].
\label{dphil}
\end{eqnarray}
The spreading is amplified by the second bracket of eq.~\mref{dphil},
which increases $|\dphil|$ with respect to the case with the aberration
and retardation ignored.\footnote{Throughout this paper the phase 
intervals that correspond to altitude differences are always defined
as the upper-altitude phase minus lower-altitude phase. 
Therefore, $\dphil$ is negative.} Similarly, for the equivalent fieldline $s$ 
on the trailing side we use eq.~\mref{phit} to give the \emph{net} change 
in phase: 
\begin{eqnarray}
\dphit & =& \phi_{\rm T}(r_2)-\phi_{\rm T}(r_1) \simeq \nonumber\\
& \simeq &\frac{3}{2}s\left[\sqrt{\frac{r_2}{\rlc}}
-\sqrt{\frac{r_1}{\rlc}}\right]- 2\left[\frac{r_2}{\rlc}-\frac{r_1}{\rlc}\right].
\label{dphit}
\end{eqnarray}
In contrast to eq.~\mref{phil}, the terms of eq.~\mref{phit}
have opposite signs which makes $|\dphit| < |\dphil|$ whenever $r_2 > r_1$
and $s \ne 0$. Because of this, the radiation on the TS `piles up' within a narrower phase range than on the LS and becomes intensified. 

If we interpret eqs.~\mref{dphil} and \mref{dphit} as component widths generated by the same fieldline $s$ on the LS and TS respectively, then we may compare them by
expanding eqs.~\mref{dphil} and \mref{dphit} about any emission radius $r$ with $r_{2}>r>r_{1}$, giving
\begin{equation}
\dphil \simeq \frac{\Delta r}{\rlc} \left(-\frac{3s}{4}\left[\frac{\rlc}{r}
\right]^{1/2} - 2\right)+O(\Delta r)^2,
\label{dphilapp}
\end{equation}
for the LS and
\begin{equation}
\dphit \simeq \frac{\Delta r}{\rlc} \left(\frac{3s}{4}\left[\frac{\rlc}{r}
\right]^{1/2} - 2\right)+O(\Delta r)^2,
\label{dphitapp}
\end{equation}
for the TS. Since both $\dphil$ and $\dphit$ are to first order
linear in $\Delta r$, 
the profile asymmetry 
does not depend on $\Delta r$
(the measure of the asymmetry is the ratio of fluxes on the TS and LS: 
$\flr = F_T/F_L=|\dphil|/|\dphit|$, see
eq.~\ref{primitive} below).
Thus, assuming the emissivity in the 
CF is roughly symmetric with respect to the $\om$-plane, one cannot change 
the relative intensity of the LS and TS by adjusting the emitter's radial 
extent $\Delta r$. 

However, this calculation fails 
in a direct vicinity of the caustic coordinates.
If the $r_{cst}$ corresponding to $s$ lies between $r_{1}$ and $r_{2}$ 
then eqs.~\mref{dphit} and \mref{dphitapp} do not give the observed TS 
component width (because $\phit$ in eqs.~\mref{dphitapp} is not a monotonic 
function of $r$ when $r$ is close to $r_{cst}$). 
In fact eqs.~\mref{dphit} and ~\mref{dphitapp} understate the component width 
because the phase shifts  above and below $r_{cst}$ partly overlap 
and cancel one another. 
 
Suppose, for a given $s$ on the TS, the radiation below $r_{cst}$ generates 
a component width (${\dphit}_{1}$) which is at least partly overlaid by the 
component width (${\dphit}_{2}$) generated above $r_{cst}$. Then the observed 
component width $\dphit$ will be the larger of ${\dphit}_{1}$ and 
${\dphit}_{2}$ and both must anyway extend to the phase $\phi_{cst}(s)$. 
To see this in eq.~\mref{dphit} we set ${\Delta r}_{i}= r_{cst} - r_{i}$ 
($i=1,2$) and expand ${\dphit}_{i}$ about $r_{cst}$ to the second order to 
give:
\begin{eqnarray}
{\dphit}_{i} & \simeq & \frac{{\Delta r}_{i}}{\rlc} 
\left(\frac{3s}{4}\left[\frac{\rlc}{r_{cst}} \right]^{1/2} - 2\right) 
+ \nonumber\\
& & +\ (-1)^{i} \frac{3}{16} s \left(\frac{\rlc}{r_{cst}}\right)^\frac{3}{2} 
\left(\frac{{\Delta r}_{i}}{\rlc}\right)^2
\label{dphitcst}
\end{eqnarray}
The first term, by definition (eq.~\ref{rcst}), is zero and the remainder has a sign depending on $i$, giving 
\begin{equation}
{\dphit}_{i} \simeq \frac{(-1)^{i}}{2}\frac{\rlc}{r_{cst}}
\left(\frac{{\Delta r}_{i}}{\rlc}\right)^2
\label{dphitcst2}
\end{equation}
This expression is independent of $s$, except indirectly through $\Delta r_{i}$.
Thus the minimum component width 
is attained at 
$\Delta r_{1}=\Delta r_2=\frac{r_{2}-r_{1}}{2}$, where the upper and lower 
phase shifts exactly overlap. 

The results of this section are likely to have important implications for 
understanding MSP 
profiles because, as mentioned earlier, these pulsars are thought to have 
deep emission zones (i.e. large $\Delta r/\rlc$) and hence caustic effects 
can be expected at a wide range of phases $\phi_{cst}$, creating narrow, 
bright components on the TS in contrast to the broad dimmer features on 
the LS. In any pulsar of given period and with fixed $r_{1}$ and $r_{2}$, 
there will be a definable \emph{caustic zone} in its profile where component 
narrowing and intensification might be expected. Its range is given by

\begin{equation}
(\phi_{cst})_{r=r_{1}}-(\phi_{cst})_{r=r_{2}}= 2\frac{\Delta r}{R_{lc}},
\label{cstrange}
\end{equation}
Note this result does not depend on the absolute value of the emission heights, and will be greatest for pulsars with narrow light cylinders.

\subsection{Caustic intensity compression}
The more strongly a given component is compressed by AR effects, the larger
will be the radio flux observed at the maximum of that component.
Thus, assuming uniform emissivity, the phase-dependent flux (profile shape)
can be estimated as:
\begin{equation}
F(\phi) \propto \frac{\Delta r}{|\dphi(\phi)|},
\label{primitive}
\end{equation}
where $|\dphi(\phi)|$ is a width of a component located at pulse phase $\phi$.
The quantity $F(\phi)$ represents the flux at the maximum of that component. 

Eq.~\mref{primitive} will be satisfactory across most of the profile 
(always including the LS), so that we are able to use the first-order 
eqs.~\mref{dphilapp} and \mref{dphitapp}. However, 
eq.~\mref{primitive} cannot be used in a direct vicinity of the
caustic coordinates. This is because our model assumes
an unphysical radiation pattern (the delta function along the electron
velocity). This leads to the infinite value of the phase-resolved
flux at $\phi = \phcst$. The infinity appears because for $\Delta r$
decreasing down to zero around $\rcst$,
the amount of radiation emitted (proportional to $\Delta r$), decreases
slower than the detection interval $\dphit \propto 1/(\Delta r)^2$
(eq.~\ref{dphitcst2}).

Although this result is formally correct,
it is unphysical because locally the radio waves are emitted
into a solid angle of finite width $\dbeam$.
Near the caustic coordinates,
that is, for $(r_1, r_2) = (\rcst - \Delta r, \rcst + \Delta r)$ and
$\Delta r \rightarrow 0$, the phase interval $\dphit$ becomes negligible
in comparison to $\dbeam$. 
In such a case, the flux observed at $\phi=\phcst$ becomes determined
by the solid angle of the elementary beam ($\propto$$\dbeam^2$),
and by $\Delta r$.
Therefore, the equation \ref{primitive}
is not applicable to the caustic points of the magnetosphere and cannot be
used to reliably predict the phase-resolved flux at the caustic phase
[$F(\phi=\phcst)$].

Such caustic effects have already been numerically
explored in discussions of the few pulsars found to have high energy 
profiles. Eq.~\mref{phcst}  is therefore  the small-angle equivalent 
of the phase at which the leading peak of the two-pole caustic 
or slot-gap model
occurs (Dyks \& Rudak 2003; Muslimov \& Harding 2004).
This leading gamma-ray peak is emitted from low altitudes on the trailing side 
of the polar tube and lags the main radio pulse by $\sim 0.1P$,
if the closest pole is viewed at a small impact angle ($\zeta \simeq
\alpha$).
In this case the phase location of this gamma-ray peak
is given by eq.~\mref{phcstnon} of the Appendix \ref{nonorth} 
for arbitrary dipole inclinations $\alpha$.

\section{The generation of asymmetric profiles.}
\label{models}
To illustrate the observable effects of AR on a millisecond pulsar we generate 
artificial profiles for a pulsar assumed to have a period of 5.25 
ms\footnote{This typical MSP period will be used throughout this paper 
to enable a comparison with the profile of J1012+5307.} and to radiate 
its emission between heights of $r_1=10^6$ cm (approximately the neutron 
star surface) and $r_2=2\cdot 10^6$ cm. Such a pulsar would have 
$R_{lc}=25 \cdot 10^6$ cm so that $\frac{r_1}{\rlc}=0.04$ and 
$\frac{r_2}{\rlc}=0.08$. 

\begin{figure}
      \includegraphics[width=0.48\textwidth]{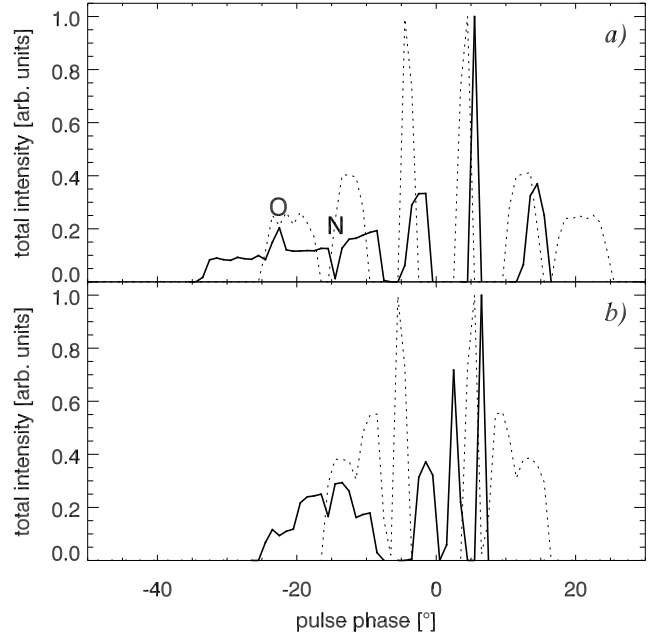}
       \caption{Modelled pulse profiles for the emission
region extending radially between $10^6$ and $2\cdot 10^6$ cm 
in a pulsar with $P=5.25$ ms and $\alpha=\zeta=90^\circ$. 
This implies a \emph{caustic zone} in the profile
ranging from $4.5^\circ$ to $9^\circ$. 
In {\bf a} the radiation is emitted from $s=0.2$, $0.6$, and $1.0$.
In {\bf b} $s=0.25$, $0.45$, and $0.65$. The profiles formed without 
AR effects are shown as dotted lines. In both cases the expected 
enhancement and narrowing of components occurs in the caustic zone 
of the TS. The result is for a fictitious emission mechanism with fixed radiative 
power per each centimeter
of electron trajectory in the IOF. Zero on the horizontal axis
corresponds to the absolute fiducial phase and the phase resolution is
$1^\circ$. }
      \label{profis}
   \end{figure}

 In Fig.~\ref{profis} an intrinsically symmetric 6-component (i.e. 3-cone) emission region is assumed 
for 2 different sets of $s$ parameters. The first (Fig.~\ref{profis}a) includes an outermost cone on $s=1$ and has a total width of $55^\circ$ ($=3\sqrt{\frac{r_2}{\rlc}}$, the opening angle of the last closed fieldline at the top of the emission zone). From eqs.~\mref{dphil} and \mref{dphit} this
gives $\dphil\simeq 15^\circ$ and $\dphit\simeq 3^\circ$, i.e.~the leadingmost
component is 5 times broader than the trailingmost. The second 
(Fig.~\ref{profis}b) 
has narrower cones and the merging of components is more comprehensive 
on the LS. In both cases, the radial extent causes the LS profiles to assume a 
`boxy' shape so that even with widely spaced cones the stretched components 
overlap and false `components' can be created (marked `O' in 
Fig.~\ref{profis}a). It can also be seen that the stretched components 
do not always fully merge with each other, so that
a single notch can appear (marked `N' in Fig.~\ref{profis}a). 
Single notches in boxy profiles were described by Cordes (1975)
for B1919$+$21. Simulated polarisation profiles (not shown) demonstrate that the overlap does not 
lead to any depolarisation, because the overlapping components
have similar polarisation angle (and our code assumes they have the same
intrinsic polarization degree). 

By contrast, the TS components are clearly narrowed and become separated. 
From eq.\mref{rcst} it follows that strong caustic effects will be observed 
on fieldlines with $s$ between 0.53 and 0.75 at phases between $4.5^\circ$ 
and $9^\circ$. This is precisely the phase range within which the strongest 
peaks occur (on $s=0.6$ in case (a) and $s=0.65$ in case (b)).
Note that the height of the strongest components is limited
by the phase resolution of our calculation (360 phase bins per period).

Note also that, although the total width of the profile remains unchanged from its CF value ($3\sqrt{\frac{r_2}{\rlc}}=55^\circ$), any central emission on $s=0$ (on a straight fieldline in the CF) would be spread over  $6^\circ$ between $4.5^\circ$ and $9^\circ$ preceding the fiducial phase.

In general, for pulsars with trailing components emitted near 
$\rcst$ the flux at the trailing edge of a profile is then likely to drop
more steeply than on the LS. 
When $\alpha\simeq90^\circ$ and $s \simeq 1$ on the trailing edge of a pulse, 
the caustic phase coincides with
the trailing edge of a profile if the emission occurs at $\rcst/(10^6\ \rm
cm) = 1$, $2$, $3.3$, $6.7$, $67$, $670$ for 
$P=1.5$, $3$, $5$, $10$, $100$, and
$1000$ ms, respectively. 
Assuming that the radio emission altitude does not exceed
a few tens of $\rns$ we expect steep trailing flanks for 
profiles of interpulsars with $P < 100$ ms.

Sharp features on the trailing side of profiles indeed
seem to be a frequent feature among MSPs (J0030+0451, Lommen et al.~2000;
J0613-0200, J1022+1001, B1855+09, B1937+21, 
Kramer et al.~1999; Stairs et al.~1999). In practice it remains difficult to disentangle the components of an MSP and hence to make estimates of $r_1$ and $r_2$. This is mainly because the true position of the fiducial phase within the pulse is unknown and the pulsar's angle of inclination, which affects our calculation through the generalised forms of eqs.~\mref{phil} and \mref{phit}  (see Appendix A), is not easy to estimate. However, as we show below, other consequences of AR may be utilised to give further clues.

%
\subsection{Position angle swing and caustic enhancement}

A simple physical description of the caustic pile-up of radiation is that
electrons at $\rcst$ for a short while move rectilinearly towards the
observer in IOF. At $r<\rcst$ their IOF trajectory is bent
in the backward/trailing
direction, whereas at $r>\rcst$ it is bent forward because of the
domination of the corotation. It is the \emph{inflection point} 
of the IOF trajectory
that is located at $(\rcst, \scst)$ [or $(\rcst, \thcst)$]. 
Thus, the region of the caustic pile-up (eq.~\ref{rcst})
is simultaneously the region of the zero-curvature of electron trajectory in
IOF, as localized by eq.~(3) in Dyks (2008). By noting that
\begin{equation}
s \simeq \frac{\theta}{(r/\rlc)^{1/2}},
\end{equation}
one can rewrite eq.~\mref{rcst} in the forms:
\begin{equation}
\frac{\rcst}{\rlc} \simeq \frac{3}{8}\thcst{\rm ,\ 
\ \ or\negthinspace\negthinspace:} \ \ \ 
\thcst \simeq \frac{3}{8} \scst^2.
\end{equation}
For a \emph{laterally}-extended emitter
the zero (or minimum) curvature corresponds to the
steepest gradient of the polarization angle swing.
The locations of sharp caustic peaks
in pulse profiles are then predicted (for a \emph{radially}-extended
emitter) to occur at the same phase, 
ie.~$\phcst = \phpa$. Indeed, eq.~\mref{phcst}
can be written as:
\begin{equation}
\phcst \simeq \frac{3}{4}\thinspace \thcst + \phf \simeq
2\frac{\rcst}{\rlc} + \phf \simeq \phpa,
\end{equation}
(cf.~eq.~16 in Dyks 2008).

The fact that 
$\phcst = \phpa$ does not always 
mean that caustic enhancement and the PA inflection point will be 
observed at the same phase 
in a given object. This is because the relativistically-delayed S-swing 
is expected
for a laterally-extended emitter, whereas the caustic enhancement
is for the radially-extended emission, in which case
the PA curve does not have to assume the S-shape. 
Some examples of the PA curves
expected for the extreme case of radial extent 
are shown in Dyks, Harding \& Rudak (2004), eg.~their fig.~5
is for $\Delta s=0$ and fig.~7 for $\Delta s\sim 0.05$. 
Rapid changes of the PA are expected at the caustic
peaks and a rather complicated PA curve, cf.~the observed optical
polarisation data of the Crab pulsar, 
S{\l}owikowska et al.~(2009). 
Moreover, for the specific 
case of  
curvature
radiation the emissivity vanishes  for zero IOF curvature, which
can produce a \emph{minimum} of flux 
at $\phcst$ in the averaged pulse profile
(
see fig.~3 in Blaskiewicz et al.~1991, and the discussion in
our Sect.~\ref{bls}). 
We anticipate the case of $\phcst = \phpa$ to be observable
for a slab-shaped emitter with the non-negligible lateral \emph{and} radial extent.

\section{Caustics and emissivity}
\label{intensity}

\subsection{Radius of curvature of electron trajectories
in the observer's frame of reference} 

The corotation increases the 
curvature of electron trajectories on the LS of the MP, whereas it decreases
the curvature on the TS (Ahmadi and Gangadhara 2002; Dyks 2008). 
The question of how this affects radio emissivity
is a complicated one, because it depends on details of the amplification
process. One one hand, one can qualitatively argue that the smaller curvature
(larger curvature radius)
makes the amplification more efficient, because the electrons following
less-curved $\vec B$-field interact with narrow beams of radiation for a longer
period of time. On the other hand, for the coherent curvature
emission from bunches of $n$ charges, the radio emissivity is proportional
to the noncoherent curvature emissivity: $F_{coh} = n^2F_{cr}$, where $F_{cr}
\propto \rhiof^{-2/3}$ is the noncoherent curvature emissivity of a single
electron in the low-frequency limit (Buschauer \& Benford 1980), 
and $\rhiof$ is the radius of curvature of electron
trajectory in the inertial observer frame (IOF). 
This holds at least as long as $n$ is independent of $\rhiof$.

  \begin{figure*}
      \includegraphics[width=0.8\textwidth]{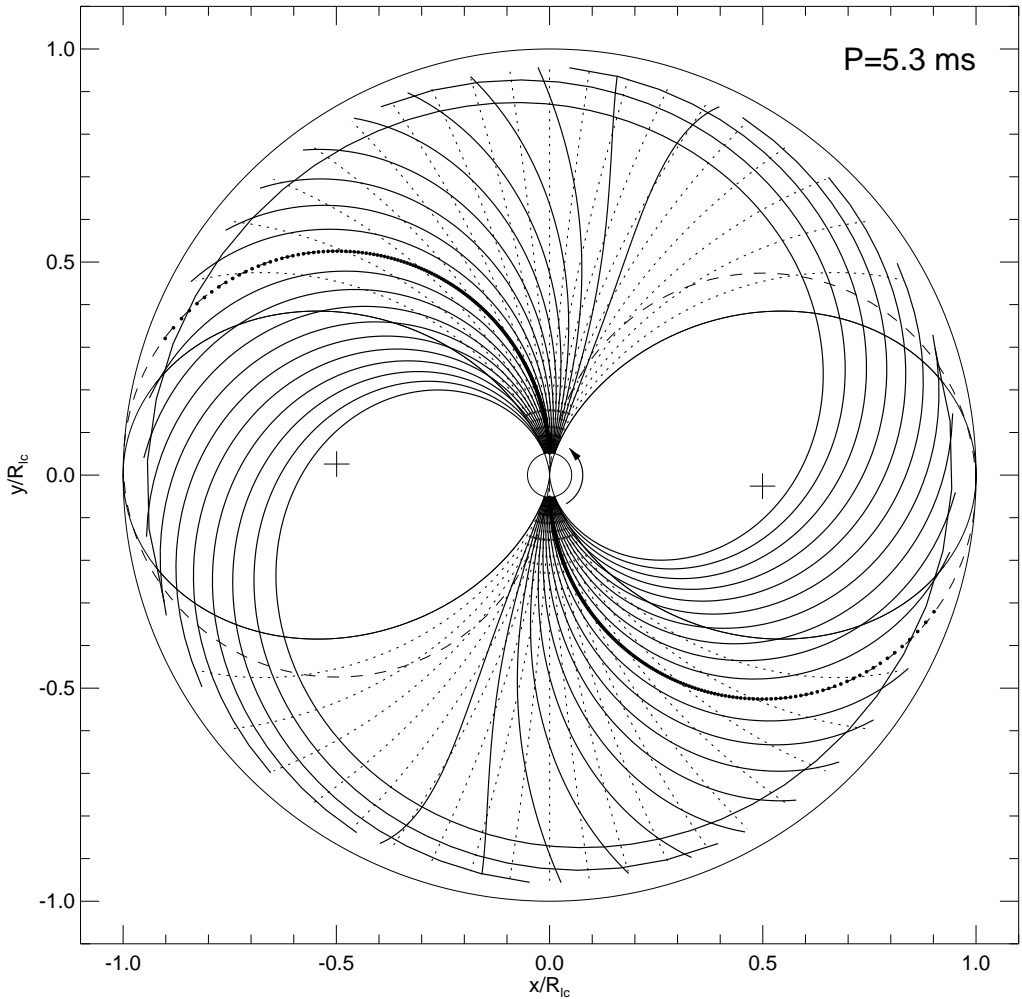}
       \caption{IOF trajectories (thick solid lines)
of electrons moving in the equatorial plane
of orthogonal pulsar ($\alpha=90^\circ$). The curves are separated by 
$\Delta s = 0.1$ and refer to electrons that left the star surface when 
the dipole axis was vertical in the figure (i.e. the fiducial instant). 
IOF trajectories for electrons moving along the dipole axis 
are marked with dots (merging into very thick solid lines)
and coincide with the analytical solution (dashed
circles of radius $\rlc/2$, 
centered at the crosses on both sides of the star; 
the centres are located at 
$x/\rlc=\pm0.5\mp(0.5\rns/\rlc)^2$ and 
$y/\rlc=\mp0.5\rns/\rlc$).
Electron trajectories in the CF ($\vec B$-field lines) are shown as dotted 
lines, except for the last open ones, which are thin solid. 
The outermost circle is the light cylinder. The figure assumes rigid
corotation of a static-shape dipole out to $0.95\rlc$. The trajectories shown 
extend from the leading edge to the 
trailing edge of the open region and have $s = 1$, $0.9$, $0.8$, ..., $0.1$, $0.0$, 
$0.1$, ..., $0.9$, $1.0$.
       }
      \label{trajs}
   \end{figure*}

For the latter mechanism our model for the pulse profile shape becomes:
\begin{equation} 
F(\phi) \propto \frac{\rhiof^{-2/3}\Delta r}{|\dphi(\phi)|}.
\label{frho}
\end{equation}
To use it efficiently, we need a convenient method for estimating $\rhiof$.
The value of $\rhiof$ can be expressed as:
\begin{equation}
\rhiof \simeq \frac{c^2}{|\vaiof|},
\label{rhiof1}
\end{equation}
where $\vaiof$ is the electron acceleration in the IOF.
In eq.~\mref{rhiof1}
the electron velocity has been approximated by the speed of light $c$,
and any acceleration parallel to the electron trajectory is thereby neglected.

The IOF acceleration can be expressed as a sum of two vectors:
\begin{equation}
\vaiof = \vec \ab + \vec \arot \simeq \frac{c^2}{\rhob}\ \hat \rhob +
2\Omega c\ \hat e_\phi,
\label{vaiof1}
\end{equation}
where $\vec \ab$ is the CF-acceleration 
caused by the curvature of magnetic lines and $\vec \arot$ is the acceleration
generated by the corotation 
(see Dyks 2008 for a simple derivation; note that
we still keep the discussion limited to
the orthogonal case of $\alpha=\zeta\simeq90^\circ$).
The symbol $\rhob$ 
denotes the radius of curvature of $\vec
B$-field lines in the CF:
\begin{equation}
\rhob \simeq \frac{4}{3}\frac{\sqrt{r\rlc}}{s},
\label{rhob}
\end{equation}
$\hat \rhob$ (with a hat) is a unit vector along that radius of curvature,
and $\hat e_\phi$ is a unit vector in the direction of the rotational
azimuth. In the small-angle approximation,
on the leading side in the plane of the rotational equator
we have $\hat \rhob \simeq \hat e_\phi$. On the trailing side
$\hat \rhob \simeq -\hat e_\phi$, so that eq.~\mref{vaiof1} becomes
one-dimensional. 


Thus, for these specific simple cases we correspondingly have:
\begin{equation}
\rhiofl \simeq  \rlc\left(\frac{3}{4}\frac{s}{\sqrt{r/\rlc}} +
2\right)^{-1} = \rlc\left(\frac{\rlc}{\rhob} 
+ 2\right)^{-1}
\label{rhiofl}
\end{equation}
(leading side),
\begin{equation}
\rhioft \simeq  \rlc\left|\frac{3}{4}\frac{s}{\sqrt{r/\rlc}} -
2\right|^{-1} = \rlc\left|\frac{\rlc}{\rhob} 
- 2\right|^{-1}
\label{rhioft}
\end{equation}
(trailing side).

Interestingly, the only difference between $\rhiof$ and the CF-curvature
radius of magnetic field lines (eq.~\ref{rhob}) is the added number `$\pm2$'
in eqs.~\mref{rhiofl} and \mref{rhioft}.
Not surprisingly, apart from the sign issue 
eq.~\mref{rhioft} is the inverse of $d\phit/dr$ (eq.~\ref{phderiv}).
This is because the very definition of the curvature radius is $\rhiof
\equiv dl/d\delta$, where $dl$ is an increment of electron trajectory
and $d\delta$ is the angular increment corresponding to $dl$.
Because of the small-angle approximation we have $dl \simeq dr$,
and because of the near-orthogonality we have $d\delta \simeq d\phi$,
ie.~a change of the direction tangent to electron trajectory
in IOF is equal to the interval of the pulse phase.
Thus, $\rhiof = dl/d\delta \simeq (d\phi/dr)^{-1}$.
For $\alpha \simeq 90^\circ$ the ratio of the IOF-curvature radii 
on the leading and trailing side can thus be conveniently
estimated with the help of eqs.~\mref{rhiofl} and \mref{rhioft}
(the model is extended to the nonorthogonal dipole inclinations
in Appendix \ref{nonorth}).

For MSPs with $P \sim$ a few ms, the value of $\rhiof$ on the trailing side of the polar cap surface
can be one order of magnitude larger than the equivalent value on the leading side. For $P=5$ ms
and $s=1$ we have $\rhiofl=5\cdot 10^6$ cm, $\rhioft=2.5\cdot10^7$ cm, whereas
the dipolar $\rhob=8\cdot 10^6$ cm. The asymmetry of $\rhiof$ about the dipole axis is
illustrated in Fig.~\ref{trajs}, where particle trajectories 
in the IOF are shown. It exhibits
trajectories from the leading edge to the trailing edge of the open 
region
[there are 21 trajectories that start from each polar cap 
with footprint parameter $s$ that goes from $1$ on the LS through $0$
to $1$ on the TS in steps of $0.1$].
The gradual increase of curvature radii towards the TS is visible.
The dipole axis is marked with merged dots (dark line).

\begin{figure}
      \includegraphics[width=0.48\textwidth]{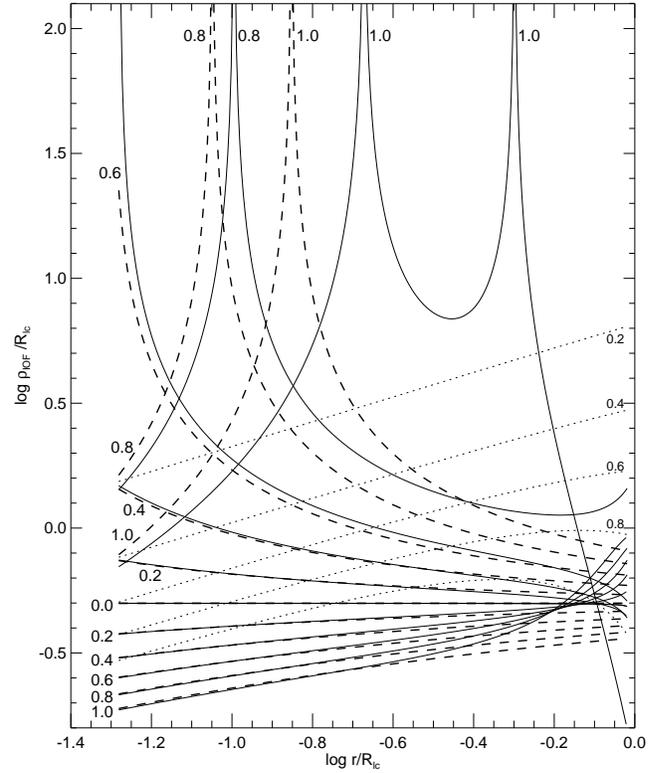}
       \caption{IOF radius of curvature calculated for selected magnetic
field lines of a pulsar with $P=5.25$ ms and $\alpha=90^\circ$. 
Solid lines present exact
numerical solution and dashed lines are for eqs.~\mref{rhiofl} and
\mref{rhioft}. The numbers give the footprint parameter $s$.
Dotted lines (the lowest one for $s=1$) present the dipolar $\rhob$. 
The horizontal marked with
`0.0' is for the dipole axis trajectory ($\rhiof = \rlc/2$)
and separates the lower curves for the LS from the upper lines for TS.
}
      \label{rhiofps}
   \end{figure}

In fact, at fixed $r$ the curvature is completely symmetric about 
the caustic trajectory, that is, about $\scst(r)$. 
This can be seen by using eqs.~\mref{rhiofl} and \mref{rhioft} to transform 
the basic equations \mref{phil} and \mref{phit}, giving
\begin{equation}
\rhiof =\frac{2r}{\left|\phi-\phi_{cst}\right|}
\label{phinew}
\end{equation}
Then, if cones are formed 
with the same IOF curvature they will be symmetric about the caustic phase 
at any given height $r$. 
 If a finite range of $r$ is considered
the cones will become ``blurred''
-- but nevertheless 
this cannot lead to the TS peaks being brought closer together.
Fig.~\ref{rhiofps} shows curvature as a function of altitude for a range 
of field line parameters (and compares this with our approximate
analytical formulae for $\rhiof$ and $\rhob$.)
This makes clear that our small angle approximation holds well 
in the region we are interested in. 

These results imply that with the knowledge of the curvature radius in the
IOF we can transfer the entire discussion of components' widths
and intensities into the IOF, without the need to refer to the effects of aberration and
retardation. Specifically, eqs.~\mref{dphil} and \mref{dphit}
are equivalent to the natural assumption that the component's width
is inversely proportional to the radius of curvature in IOF:
\begin{equation}
\dphi \simeq \frac{\Delta r}{\rhiof}.
\end{equation}
This can be immediately proved by comparison of the approximate
eqs.~\mref{dphilapp} and \mref{dphitapp} with the formulae for $\rhiof$
(eqs.~\ref{rhiofl} and \ref{rhioft}).
\emph{The caustic effects of AR can thus be seen as simply the
consequence of an increase of the curvature radius in IOF.}

Our flux model of eq.~\mref{frho} can then be written
\begin{equation}
F(\phi) \propto \frac{\rhiof^{a}\Delta r}{\Delta r/\rhiof}=\rhiof^{1-a}.
\label{frho2}
\end{equation}
For the non-coherent, low-frequency curvature radiation, the
exponent $a$ is equal to $-2/3$, and the equation \mref{frho2} 
just reduces to $F(\phi) \propto \rhiof^{1/3}$, which 
suggests a weak amplification on the TS. However, as we shall see in the 
next subsection, more physical considerations are likely to reverse this 
effect.

As a consequence, the TS of the profile is subject to two potentially competing effects, both dependent on trajectory curvature and hence both having maximum effect around the caustic zone of the profile. On the one hand, the AR bunches the emission of a single fieldline into a narrow phase. On the other, if the coherent radio emissivity decreases with increasing $\rhiof$ (as the model of eq.~\ref{frho} assumes), this may counteract the caustic amplification.

\subsection{Giant pulses at the caustic singularity?}

It is also interesting to consider the implications of caustic enhancement 
for single pulses. The quasi-rectilinear motion of electrons in the
caustic regions of magnetosphere makes favourable conditions for
coherent amplification, because the interaction between
the emitted radiation and particles can last for a longer period of time than
in the case of a strongly-curved trajectory.
We speculate that this may give rise to `giant' pulses, whose 
phase location is determined by eq.~\ref{phcst} (see Appendix A for the 
non-orthogonal version). At the same time, however, the reduced curvature decreases the
emissivity and frequency of the curvature radiation. To overbalance this
effect, the efficiency of the amplification process would have to grow 
with decreasing curvature sufficiently quickly. The energy of electrons
has also to be larger for the curvature radiation to extend up to the
observed band.

\subsection{Rotation and the profile flux balance}
\label{bls}

Eq.~\mref{frho2} enables us to consider the profile asymmetry
in more realistic cases, differing from the basic cases of Fig.1 where $\Delta r$ was held constant across the profiles and only AR effects were taken into account. Let us assume an emission region with the quantity $\Delta r$ symmetric with respect to the main meridian so that the only asymmetric quantity 
left is $\rhiof$.  
We also allow for a non-negligible range of the footprint parameter 
$\Delta s$
associated with a given component because it is easy to model numerically 
and can be expected in realistic cases. The profile asymmetry is also 
determined by the value of the exponent $a$
that describes how the intrinsic coherent emissivity depends on $\rhiof$.
In the discussion below we assume $a=-2/3$ for definiteness
[non-coherent curvature radiation (CR) in the low-frequency limit]. 

The effect of non-zero $\Delta s$ can be modelled by modifying 
eq.~\mref{frho2} so that
\begin{equation}
F(\phi) \propto \frac{\rhiof^{a}\Delta r}{\Delta r/\rhiof+\frac{3}{2}\sqrt{\frac{r}{R_{lc}}}\Delta s}.
\label{frho3}
\end{equation}

\begin{figure}
      \includegraphics[width=0.49\textwidth]{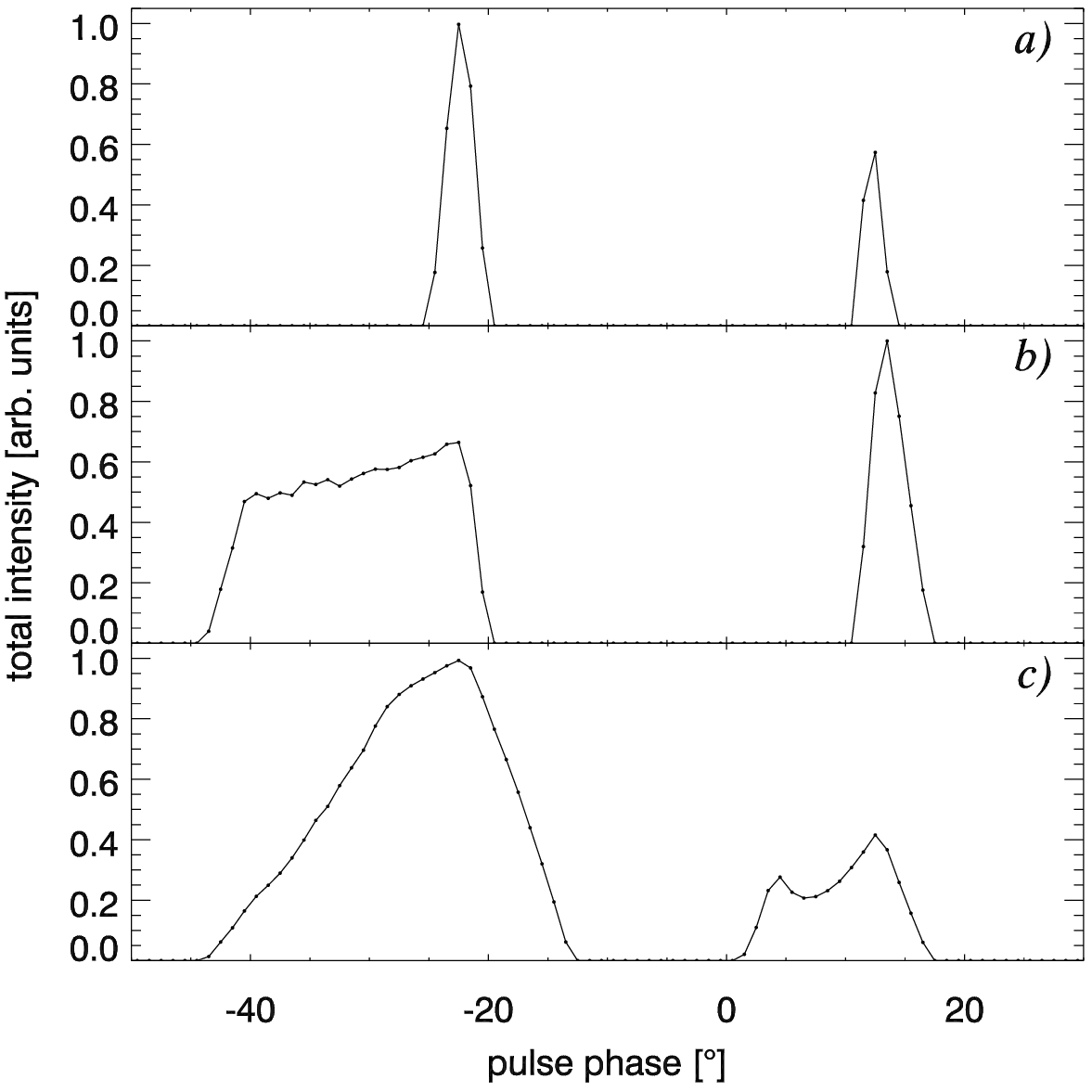}
       \caption{Simulated pulse profiles that show how the rotational asymmetry
                is affected by changes of $\Delta r =
\left<r_1,r_2\right>$ and $\Delta s=\left<s_{\rm min},s_{\rm
                max}\right>$.
                In all figures the intrinsic emissivity is proportional to 
                $\rhiof^{-2/3}$ (non-coherent CR).
                In {\bf a)} $\Delta s =\left<0.9,1\right>$, and 
               $\Delta r/\rlc=\left<0.04, 0.05\right>$.
                In {\bf b)} $r_2$ is increased to $0.12\rlc$.
                In {\bf c)} $r_2=0.12\rlc$ and $s_{\rm min} = 0.5$.
       }
      \label{dsn}
   \end{figure}

In the case of negligible $\Delta r$ (ie.~for $\Delta r/\rhiof \ll \frac{3}{2}\sqrt{\frac{r}{R_{lc}}}\Delta s$
in eq.~\ref{frho3}) we have $F\propto \rhiof^{-2/3}$, ie.~the components
on TS of the profile are dimmer than those on the LS. This can be seen
in Fig.~\ref{dsn}a which is calculated for
$\Delta r=\left<r_1,
r_2\right>=\left<0.04,0.05\right>\negthinspace\rlc$
and $\Delta s= \left<s_{\rm min},s_{\rm
                max}\right>=\left<0.9,1\right>$.
Fig.~3 in Blaskiewicz et al.~(1991) presents a similar case for a region
completely filling in the open region [$\Delta s=\left<0,1\right>$].

In Fig.~\ref{dsn}b the upper radial distance $r_2$
is increased to $0.12\rlc$. This makes the caustic term $\Delta r/\rhiof$
in eq.~\mref{frho2} more important so that we have $F\propto
\rhiof^{-2/3}\rhiof$ and the TS component becomes brighter
[the component's `brightness' 
is understood as the flux at the maximum of the component, not the
phase-integrated flux].
For $\Delta s = \left<0.5,1\right>$ and $r_2=0.12\rlc$ (Fig.~\ref{dsn}c),
the decline of the intrinsic emissivity 
($F_{\rm coh}\propto \rhiof^{-2/3}$) again dominates
over the caustic amplification and the LS component becomes brighter.
In addition to the eq.~\mref{frho3} this change of $\flr$ can be understood
as follows. The peak flux on the LS is increased because the emission from 
adjacent $s$ overlaps in phase.
On the TS the caustic peaks associated with different $s$
do not occur at the same phase. The larger value of $\Delta s$ 
makes the components on the TS wider, but their peak flux is not
enhanced considerably.
The influence of $\Delta r$ and $\Delta s$ on the pulse shape can therefore
be summarized in the following two points: 1) the increase of $\Delta r$ makes
the LS components wider and the TS components stronger (larger peak flux,
Fig.~\ref{dsn}b);
2) the increase of $\Delta s$ works in just the opposite way: 
it makes the LS components stronger and
the TS components wider (Fig.~\ref{dsn}c).\\ 
When the emissivity is independent of $\rhiof$ ($a=0$) it is the TS that is
\emph{always} brighter, with the flux ratio determined by the relative values
of $\Delta r$ and the intrinsic 
width $\dphintr = (3/2)\sqrt{r/\rlc}\Delta s$. 

We therefore conclude that the caustic amplification
($\propto \rhiof$) makes the TS brighter only if the intrinsic emissivity
does not decrease with increasing $\rhiof$ too fast. 
For the low-frequency CR ($a=-2/3$) the caustic amplification 
can still dominate over the
intrinsic flux decline (as in our 
Fig.~\ref{dsn}b)
but this requires special conditions to be fulfilled (small $\Delta s$,
large $\Delta r$). 

For large $\Delta r$ and several emission cones, 
the above conclusions are furthermore 
complicated by the possible
overlap of different components on the LS, which can additionally
increase the peak flux on the LS.
In the simulated profile of Fig.~\ref{profis}b the two maxima
at $\phi=-18^\circ$ and $-14^\circ$ are created by an overlap of
emission pattern from three cones.
Such an inter-component overlapping is not included 
in eqs.~\mref{frho2} and \mref{frho3}.

We then find that it is possible (and not unlikely)
for the LS to become brighter purely because of the intrinsic dependence
of emission mechanism on $\rhiof$. 
For $a=-2/3$ the TS can be enhanced only under special 
circumstances (small $\Delta s$, not overlapping components).
Narrow and strong trailing components seem to be a frequent phenomenon
among MSPs (Kramer et al.~1998; 1999). Moreover, at least 
some exceptions (with apparently brighter LS) seem to be
created by the scattering phenomenon (see eg.~fig.~8 in Kramer et al.~1999). 
This suggests that $-2/3<a\la0$, provided we are not biased
by unrecognized absorption effects.
The results of this section also apply for the models of high-energy
emission (even for the polar cap model if applied for pulsars with 
millisecond periods). In the outer gap and slot gap models, the
peaks in the gamma-ray profiles are generated mostly by the caustic effects.
They are then formed in the region where the IOF radius of curvature
is greatly increased in comparison to the CF value. It is then absolutely
necessary to use the IOF curvature radius $\rhiof$
to obtain reliable pulse profiles in the numerical simulations
of the CR.


\section{Pulsars with phase-shifted cones}
\label{curvasym}

Gangadhara \& Gupta (2001, hereafter GG01),
noted that profiles of selected pulsars have outer cones more shifted
towards early phase than inner cones. Krzeszowski et al.~(2009)
find at least a tendency towards this effect in most pulsars in their
sample, but they also find some counterexamples (e.g.~B1831$-$04).
The forward shift of outer cones was interpreted by GG01 
in terms of a lower emission altitude for an inner cone.
The detection of such a low-altitude inner cone is delayed
by a decrease in the aberration angle and the increased 
propagation time. 
This interpretation may be called the
`cone-altitude effect'. It implies a structured emission region with two rings of different size
located at two different altitudes (and with a core emission region near the
surface).

Because such a structure has never received a physical explanation
it is worth considering whether an alternative scenario exists in which
the cones can be actually located at the same altitude,
but intrinsically decentered in the CF to produce the forward shift of the cones.
A possible natural argument for this is the shift of 
the pair-production symmetry point towards the TS.

\begin{figure}
 \includegraphics[width=0.48\textwidth]{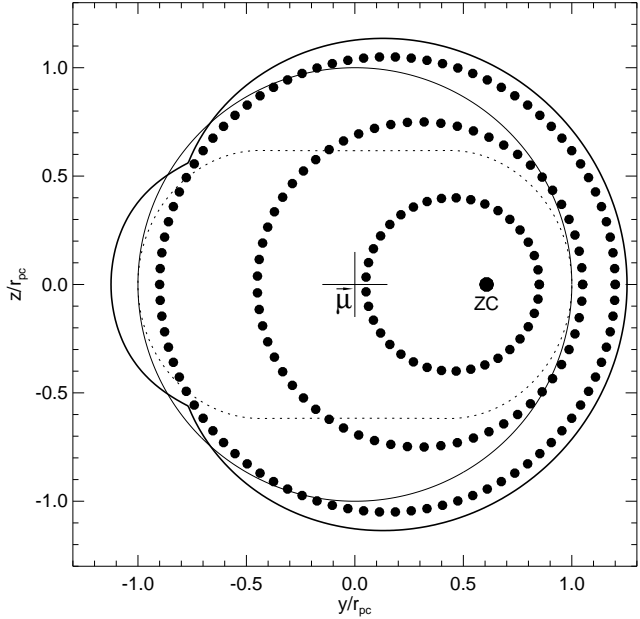}
      \caption{Location of the zero-curvature point (bullet marked `ZC')
at the polar cap of a pulsar with $\alpha=90^\circ$, $P=5.25$ ms, 
$\rns=1.3\cdot 10^6$ cm. The curvature of electron trajectories in IOF 
(thereby the curvature radiation) vanishes at the point where $s=0.61$.
Curved CF trajectories of non-curvature photons that are emitted there 
coincide with the local $\vec B$-field line in the CF. Thick solid and dotted
lines mark the polar cap edge for the retarded and static-shape dipoles.
Thin solid circle marks the standard circular polar cap (formally for
$\alpha=0^\circ$). The decentered circles of dots show possible
locations of emission cones affected by the ZC point.}
      \label{caps}
  \end{figure}

As shown in Fig.~\ref{caps}, for $P=5.25$ ms and $r=1.3\cdot 10^6$ cm, 
the zero curvature of electron trajectories in the IOF is expected
at $\scst=0.61$ on the TS of the open region, 
ie.~at a phase $\phi \simeq \phi_{\rm MP1} 
+ 0.8W_{\rm MP}$, where $W_{\rm MP}$ is the profile width
and $\phi_{\rm MP1}$ is the location of the first (leadingmost) 
component.\footnote{In the case of a rotationally-distorted vacuum dipole
the phase is equal to $\phcst(\rns) = \phi_{\rm MP1} + 0.68W_{\rm MP}$,
because of the boundary of the open region is shifted toward the TS.}
Conversely, any non-curvature
photons emitted by the electrons (moving 
along the $\vec B$-field line with $s=0.61$)
propagate in the CF along a
\emph{curved} path locally coincident with this $\vec B$-field line 
and are least likely to participate in any pair-creation cascade 
(see fig.~7 in Dyks \& Rudak 2002). 
The point of zero IOF curvature is therefore also the location 
of vanishing (or minimum) pair production.
The density distribution of the created $e_\pm$-pairs 
is then expected to be symmetrical around 
the ZC point located clearly on the TS of the open 
region.
The same symmetry (centered at the zero-curvature coordinates)
can be imposed on the accelerating electric field
if it is strongly screened by the pairs. 

On the other hand, the general geometry of the open region
implies the dipole-centered symmetry.
If the outer boundary $s=1$ 
has indeed the same electric potential
everywhere, the (unscreened) electric field should roughly follow 
the symmetry of the
open dipolar fieldlines (which are centered on the dipole axis, not at the 
lowest-curvature
IOF-trajectory). It is therefore conceivable that any 
conal emission structures within the open region will be affected 
by both symmetries and 
may assume the `intermediate-case' geometry shown in Fig.~\ref{caps}.
\emph{This effect can mimic the cone-altitude effect even if
all the cones originate at the same altitude.} 

\section{Application to the main pulse of J1012$+$5307}
\label{MSP}

\subsection{Component widths}
\label{widthsapp}

\begin{figure}
      \includegraphics[width=0.48\textwidth]{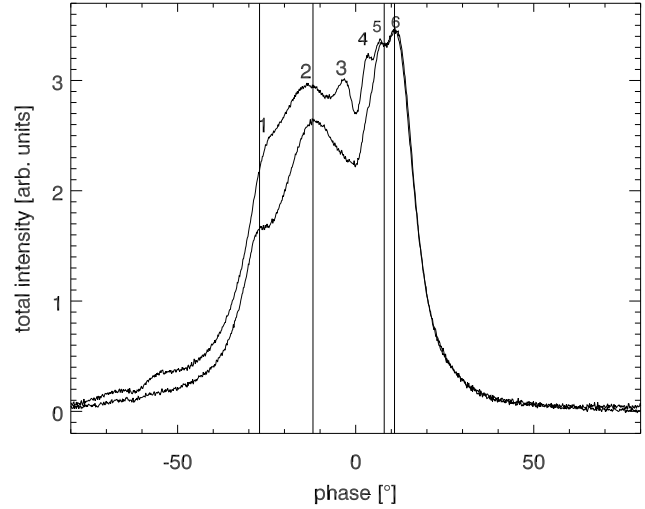}
       \caption{The main pulse of J1012+5307. Radial extent of the radio emission implies different 
widths of components on the leading and trailing side of the MP.
The two pairs of vertical lines present the components' widths expected for
low-altitude emission (between $r_1 = 10^6$ and $r_2 = 2.3\cdot10^6$ cm) 
from the last open field lines
($s=1$).
       }
      \label{widfig}
   \end{figure}

Fig.~\ref{widfig} presents the main pulse (MP) of a $5.26$ ms pulsar
\jtt observed at $0.82$ GHz (top) and $1.4$ MHz (bottom) 
with the Green Bank Telescope. The bandwidths were $64$ MHz 
at both frequencies and the total integration time of $\sim$$15$
hours spanned the period between July 2004 and March 2007.
This pulsar was chosen because it exhibits an interpulse (not shown), strongly suggesting a near-orthogonal orientation of rotational and dipole axis.
The main pulse profile exhibits asymmetry which looks consistent
with the implications of eqs.~\mref{dphil} and \mref{dphit}:
the components on the leading side are very broad and strongly merged with each
other, whereas the components on the trailing side, despite being located
much closer to each other, can still be identified through the noticeable
minima between them. The trailing components must then be much 
narrower than the leading ones.
Because of the blending the exact width of the components cannot be
determined. However, because the components are nevertheless identifiable, 
we assume that their widths are similar to the components' separation.
  
For the leadingmost component of the MP we 
use the width of $\dphil \simeq -15^\circ$ which is marked by
the leftmost pair of vertical solid lines in Fig.~\ref{widfig}.
The width can be used to estimate the corresponding range of radial distance
$\drl$.
From eq.~\mref{dphil} we have:
\begin{equation}
\frac{r_2}{\rlc} 
\simeq \left(\sqrt{\frac{9}{64}s^2 - \frac{1}{2}
\left[\dphil - \frac{3}{2}s\sqrt{\frac{r_1}{\rlc}} - 
2\frac{r_1}{\rlc}\right]} -\frac{3}{8} s \right)^2.
\label{rld}
\end{equation}
For $r_1 = 10^6$ cm and $s = 1$ we get $r_2 \simeq 2.3\cdot 10^6$
and $\drl=r_2-r_1\simeq 1.3\cdot 10^6$ cm. Thus, the width of components on the
leading side of the MP implies the radial extent of the order of the
NS radius $\rns$.

Assuming approximate symmetry of the outer cone with respect
to the $\om$-plane in the CF, 
one can use the derived value of $\drl$ 
in eq.~\mref{dphit} to estimate the width of the components on the TS
of the MP and thereby to test the reasoning. For $r_1=10^6$ cm and $r_2 =
2.3\cdot 10^6$ cm eq.~\mref{dphit} gives $\dphit \simeq 2.9^\circ$, which is
marked in Fig.~\ref{widfig} by the pair of the vertical solid lines on the right-hand
side. The 
width-scale of the leading and trailing components
is then consistent with the radio emitter's radial extent of
the order of $\sim 10^6$ cm.

\subsection{Component heights}

However, the interpretation shown in Fig.~6 cannot be made consistent with
the symmetry properties of $\rhiof$. According to eq.~\mref{phinew}, 
the IOF radius of curvature is symmetrical
with respect to the caustic phase.
Thus, both the caustic effects and $\rhiof$-driven emissivity
should be symmetrical with respect to $\phcst$. 
For $r < 0.14\rlc$ the caustic phase always precedes the
trailingmost component (if the latter is assumed to have $s=1$).
The height of components, ie.~the flux $F(\phi)$, is then expected to fall 
off symmetrically on both sides of some phase located within the pulse 
window. 
In the main pulse of \jtt the flux increases monotonically towards 
its extreme trailing edge
which can be interpreted in two ways: 1) the profile is shaped by physical 
effects independent of $\rhiof$; 
2) the emission region
is not centered at the dipole axis, but is located fully on the LS
of the caustic coordinates.

\begin{figure}
      \includegraphics[width=0.48\textwidth]{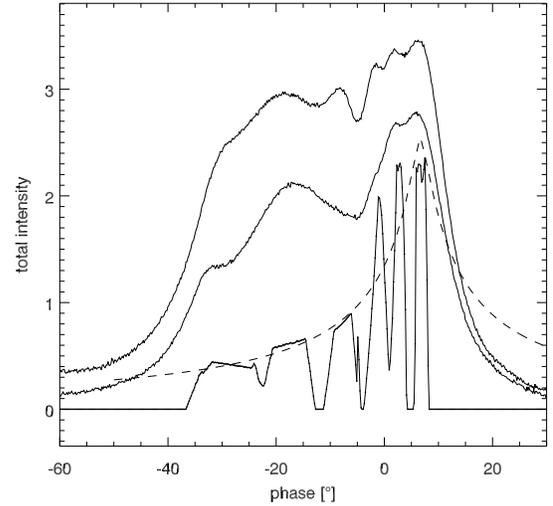}
       \caption{Numerically-calculated pulse shape (bottom curve)
drawn below the $\jtt$ data. This uniform-emissivity model has $r_1=10^6$ cm, 
$r_2=2\cdot10^6$ cm, $\Delta s=0.1$, and $\alpha=\zeta=90^\circ$.
Dashed curve presents eq.~\mref{frho3} with $a=0$ and assumes that
$\phcst=7^\circ$.
       }
      \label{fixalt}
   \end{figure}

In the second case, $\phcst$ needs to be located somewhere
on the right-hand side of the trailingmost component.
Since no observed features can be noticed there, and the available model
parameters allow for some fit ambiguity, in Fig.~\ref{fixalt} 
we present just a sample profile calculated for $\phcst$ located
at the phase of the trailingmost component. The modelled profile
(bottom solid line) has been obtained for $r_1 = 10^6$, $r_2 = 2 \cdot 10^6$
cm, $\Delta s=0.1$, $\alpha=\zeta=90^\circ$ and for uniform emissivity
per centimeter of IOF trajectory.
The component on the extreme right-hand side is located right at the
caustic phase $\phcst \simeq 7^\circ$ (for $\rav = 1.5 \cdot 10^6$ cm). 
Its width and height are therefore mostly determined by the
magnitude of $\Delta s$. The flux on the LS is dominated 
by the effects of $\rhiof$. In these two limits the flux
is well approximated with the analytical model \mref{frho3} (dashed line,
symmetric around $\phcst$). In the intermediate region [phase range
$(-5^\circ,5^\circ$)] the agreement is worse.
The values of $s$ parameter have been selected for each component
to match the observed pulse phase and vary from $s=1$ for the
component on the leading edge, through $s=0$ near the central minimum,
to the caustic value of $s=0.6$ for the trailingmost component.

The numerical profile exhibits the general observed properties, 
ie.~the monotonic increase of flux towards the TS and 
the narrower components on the TS. The LS/TS flux ratio could be made more
consistent with data by placing $\phcst$ far on the trailing side.
However, such fit would not be unique, because
the ratio can also be adjusted by changing $\Delta s$ and $\Delta r$.
The difficulty in establishing $\phcst$ 
makes any
multidimensional fitting procedure somewhat ambiguous and discourages us
from forcing the model to closely match the data.
The LS/TS asymmetry can also be probably 
decreased by increasing 
the impact angle $\beta = \zeta-\alpha$.
Moreover, the observed profiles seem to be affected 
by the overlapping of neighbouring components.
This cannot be easily modelled with our analytical formulae 
nor with the numerical code, which creates pronounced spurious components
caused by the sharp boundaries of the emission region at $r_1$ and $r_2$.
To avoid them,
the code would have to be supplied with additional parameters
that control the radial profile of emissivity. 

The CR is physically better justified
than the uniform emissivity model shown in Fig.~\ref{fixalt},
and it has been shown to produce enigmatic bifurcated components
in the profile of J1012$+$5307 (Dyks, Rudak \& Demorest 2009). 
In the case of the CR,
a minimum of flux is expected at $\phcst$. This suggests that the caustic
phase be associated with the central minimum of the MP.
We do not model this case, however, because the observed MP is not
symmetric about this minimum (whereas the modelled profile would have been
symmetric).

An alternative way to successfully model the profile is the cone-altitude
approach of GG01, which assumes a set of the dipole-axis-centered rings
of emission in the CF. Assuming that the near-surface emission
within the main meridian is detected at the central minimum,
the observed profiles can be reproduced if the rings are located at 
$s=0.53$, $0.7$, and $0.84$, and the average emission radius is
equal to $1$, $1.7$, and $2.5\cdot 10^6$ cm, respectively.

Apart from \jttns, a similar systematic increase of flux towards the TS
is exhibited by the central structure in the 438 MHz profile of
J0437$-$4715 (Navarro et al.~1997).

\section{Conclusions}
\label{conc}

We find that the shape of a profile is affected by 
two opposing effects of rotation:
1) the caustic enhancement of the TS, and 2) the increase of $\rhiof$ on the
TS which decreases the centripetal acceleration and can thereby weaken the
emissivity. The first can be viewed as a direct
result of the straightening of an electron trajectory in the IOF, because
the radiation emitted within $\Delta r$ 
is received within $\dphi \simeq \Delta r/\rhiof$.
The radiation that constitutes the caustic peaks is emitted from
the inflection points of IOF trajectories (or from regions of minimum
IOF curvature in a non-orthogonal case).
Formally, the caustic peaks are expected to occur at the phase
of the center of the relativistically-shifted S-curve of the polarisation angle.
The caustic phase, if calculated for the maximum $s=1$, becomes
the `maximum possible detection' phase for tangent-to-$\vec B$ radiation
within the open region. Fast-rotating and highly-inclined pulsars
are then expected to have pulse profiles with sharply defined 
(distinctively marked) trailing sides. On the LS, 
AR causes radiation from the outflowing electrons 
to be detected at a phase that monotonically increases with the emission
altitude.
The LS of  the profile
is thus expected to grow only gradually, creating a leading wing
(the extent of which in practice is likely determined 
by the cooling timescale of the
electrons).

Whether the leading or trailing side of the profile is the more enhanced 
by rotation 
is likely to be determined by the intrinsic dependence of the coherent 
emission 
mechanism on $\rhiof$.
In the case of the CR the emissivity is entirely driven 
by the macroscopic
centripetal acceleration and we may expect a minimum in the zero 
curvature region. Otherwise, a caustic peak will dominate the profile 
(cf.~Fig.~\ref{profis}b with \ref{dsn}c). Statistical analysis
of millisecond profiles, could therefore provide limits
on the possible value of the exponent $a$ in the relation 
$F_{coh}\propto \rhiof^a$. Inspection of published profiles
suggests that $a\la 0$ (as opposed to $a=-2/3$), although the number of 
good-quality millisecond
profiles is very limited, and the interpretation of profiles in terms
of a simple sum of emission components ignores possible
absorption effects.

According to popular electrodynamic
assumptions, the accelerating electric field is symmetrical with respect to
the main meridian. However, 
the IOF curvature radii are intrinsically symmetric about 
the caustic phase at a given height (and asymmetric with respect to the
dipole axis), so circular cones centered on this phase 
are theoretically likely.
Because of the IOF-curvature asymmetry within the open region of 
fast rotating MSPs, the traditional, dipole-axis-centered  
nest of cones may be considered 
less natural than the decentered geometry of Fig.~\ref{caps}.
In the extreme case of the fastest known MSPs ($P\sim 1.5$ ms), 
the caustic phase occurs at the outermost trailing edge of the polar cap, 
so it becomes difficult to generate a conal structure at all
(possibly yielding the simplicity of the MP of B1937$+$21).

The main pulse of \jtt 
exhibits asymmetry of flux and width which can be interpreted
in terms of the corotational effects within a radially-extended emission
region. The observed width of the main pulse suggests that 
the outermost components have $s \simeq 1$. This is roughly consistent 
with the widths of the outermost components 
if $r_1 \sim \rns$ and $\Delta r \sim \rns$.
However, based solely on the symmetry of $\rhiof$, 
the profile should then be symmetrical with respect to the caustic phase
located somewhere within the main pulse window, which is not the case.
This implies that either the profile is dominated by unknown 
$\rhiof$-independent effects, or the emission region 
is located fully on the LS of the caustic coordinates.
In the uniform-altitude scenario (same $r_1$ and $r_2$ for all components)
this requires the rings/cones to be located asymmetrically 
with respect to the main meridian plane, but the location is not uniquely
constrained because of the degeneracy of parameters. 
In the alternative cone-altitude scenario, the asymmetry with respect to 
$\scst$ can still be achieved by a region that is symmetrical with respect
to the main meridian in th CF. However, it is hard physically to justify the
cone-altitude geometry. Neither of the scenarios 
is naturally consistent with the observed width of the main pulse, which
is the same as the opening angle of the near-surface polar beam.

A possible solution to this problem is a model in which the
polar tube at low altitudes is fully filled with the radio emission,
and the minima observed in the MP result from absorption effects.
The absorption could be provided by dense, opaque, radially-extended 
plasma columns in the polar tube.
In this model the MP is not regarded as a sum of emission components.
Rather, it is a single, box-like feature, with the minima 
of different width indented by the plasma columns. 
We expect that the minima will have essentially the same widths
as the emission components in the emissive version of the model
presented in Section 2. That is, the minima on the LS/TS will
correspondingly be broad/narrow. Then, 
if the absorption occured
(on average) above $0.14\rlc$, it would be possible to avoid 
the problem of profile symmetry around $\phcst$. 
Such an absorption-based scenario is supported by the box-like look 
of the MP of J1012$+$5307, and presents a promising subject for future
study.

\section*{Acknowledgements}

GAEW thanks the University
of Sussex for a Visiting Research Fellowship.
This paper was supported by the grant N203 387737
of the Ministry of Science and Higher Education.

\appendix

\section{Extension of the model to non-orthogonal pulsars}
\label{nonorth}

For $\alpha \ne 90^\circ$ the rotationally-induced
acceleration becomes $\vec \arot\simeq 2\Omega c \sin\alpha \hat e_\phi$ 
(Dyks 2008)
so that the IOF radius of curvature becomes
\begin{equation}
\rhiof \simeq  \rlc\left|\frac{3}{4}\frac{s}{\sqrt{r/\rlc}} \pm
2\sin\alpha\right|^{-1},
\label{rhioflt}
\end{equation}
where the plus sign corresponds to the LS, whereas
the minus is for the TS.

Near the star surface
the equations hold with accuracy of $\sim 1\%$ (for $P\simeq 1$ s and
$\alpha$ as small as  $20^\circ$). For
$P\simeq 5$ ms (and $\alpha=20^\circ$) the accuracy is degraded
down to typically $\sim10\%$ and can be worse near 
the locations of zero curvature (cf.~Fig.~\ref{rhiofps} for
$\alpha=90^\circ$).

If $\zeta \simeq \alpha$ it is also possible to extend 
the applicability of eqs.~\mref{phil}-\mref{dphit}
to the non-orthogonal case of $\alpha\ne 90^\circ$.
To do this the square root terms in eqs.~\mref{phil}-\mref{dphit}
(but \emph{not} the linear terms)
need to be multiplied by $1/\sin\alpha$. This is because the
`not a great circle effect' expands the profile width while
leaving the linear AR shift intact.

For example, the detection phase for the TS radiation becomes:
\begin{equation}
\phi_{\rm T} \simeq \frac{3}{2}\frac{s}{\sin\alpha}\sqrt{\frac{r}{\rlc}} 
- 2\frac{r}{\rlc} +
\phf,
\label{phitnon}
\end{equation}
so that the caustic amplification determined by the condition 
$d\phit/dr=0$ occurs for radiation emitted from:
\begin{equation}
\frac{r}{\rlc} \simeq \frac{9}{64}\frac{s^2}{\sin^2\alpha}.
\label{rcstnon}
\end{equation}
Thus, for a given $r$ the caustic effects occur more in the central parts
of the profile (smaller $s$) than in the orthogonal case, whereas 
for fixed $s$ the caustic enhancement occurs at larger $r$.
The amplification is observed at the phase:
\begin{equation}
\phcst \equiv \phi_{\rm max} \simeq \frac{9}{32}\frac{s^2}{\sin^2\alpha} + \phf = 
16^\circ\kern-1.25mm.11 \frac{s^2}{\sin^2\alpha} + \phf,
\label{phcstnon}
\end{equation}
ie.~the caustic peaks occur further behind the MP than in the orthogonal
case.

Though not related to the LS/TS asymmetry, it is also easy
to find the IOF radius of curvature in the main meridian plane, where
$\hat \rhob \perp \hat e_\phi$. Using eqs.~\mref{rhiof1}, \mref{vaiof1},
and the Pythagorean theorem we get
\begin{equation}
\rhiof \simeq  \rlc\left(\frac{9}{16}\frac{s^2}{r/\rlc} + 4\sin^2\alpha
\right)^{-1/2}.
\label{rhiofsmm}
\end{equation} 

\bsp

\label{lastpage}

\end{document}